\documentclass[twocolumn,showpacs,aps,prd,superscriptaddress,floatfix]{revtex4-2}
\usepackage{graphicx}  
\usepackage{dcolumn}   
\usepackage{bm}        
\usepackage{amssymb}   
\usepackage{amsfonts}  
\usepackage{hyperref}
\usepackage{amsmath}
\usepackage{amsthm}
\usepackage{mathtools}
\usepackage{xspace}
\usepackage{gensymb}
\usepackage{mathrsfs}

\def \gev  {\mbox{GeV}}

\def \gevcc{\mbox{GeV/$c^2$}}

\def \mevcc{\mbox{MeV/$c^2$}}

\def \gev  {\mbox{GeV}}

\def \gevcc{\mbox{GeV/$c^2$}}

\def \mevcc{\mbox{MeV/$c^2$}}

\def\simge{\mathrel{
   \rlap{\raise 0.511ex \hbox{$>$}}{\lower 0.511ex \hbox{$\sim$}}}}
\def\simle{\mathrel{
   \rlap{\raise 0.511ex \hbox{$<$}}{\lower 0.511ex \hbox{$\sim$}}}}

\begin{document}
\title{\bf Search for di-photon decays of an axion-like particle in radiative  \boldmath{$J/\psi$} decays}

\author{
  \begin{small}
    \begin{center}
    M.~Ablikim$^{1}$, M.~N.~Achasov$^{4,b}$, P.~Adlarson$^{75}$, X.~C.~Ai$^{80}$, R.~Aliberti$^{35}$, A.~Amoroso$^{74A,74C}$, M.~R.~An$^{39}$, Q.~An$^{71,58}$, Y.~Bai$^{57}$, O.~Bakina$^{36}$
% ...
         \\
      \vspace{0.2cm}
      (BESIII Collaboration)\\
      \vspace{0.2cm} {\it
    $^{1}$ Institute of High Energy Physics, Beijing 100049, People's Republic of China\\
% ...
$^{4}$ Budker Institute of Nuclear Physics SB RAS (BINP), Novosibirsk 630090, Russia\\
% ...
$^{35}$ Johannes Gutenberg University of Mainz, Johann-Joachim-Becher-Weg 45, D-55099 Mainz, Germany\\
$^{36}$ Joint Institute for Nuclear Research, 141980 Dubna, Moscow region, Russia\\
% ...
$^{39}$ Liaoning Normal University, Dalian 116029, People's Republic of China\\
% ...
$^{57}$ Southeast University, Nanjing 211100, People's Republic of China\\
$^{58}$ State Key Laboratory of Particle Detection and Electronics, Beijing 100049, Hefei 230026, People's Republic of China\\
% ...
$^{71}$ University of Science and Technology of China, Hefei 230026, People's Republic of China\\
% ...
$^{74}$ University of Turin and INFN, (A)University of Turin, I-10125, Turin, Italy; (B)University of Eastern Piedmont, I-15121, Alessandria, Italy; (C)INFN, I-10125, Turin, Italy\\
$^{75}$ Uppsala University, Box 516, SE-75120 Uppsala, Sweden\\
% ...
$^{80}$ Zhengzhou University, Zhengzhou 450001, People's Republic of China\\
\vspace{0.2cm}
% ...
$^{b}$ Also at the Novosibirsk State University, Novosibirsk, 630090, Russia\\
% ...
      }\end{center}
    \vspace{0.4cm}
  \end{small}
}

\begin{abstract}
We search for the di-photon decay of a light pseudoscalar axion-like
particle, $a$, in radiative $J/\psi$ decays, using 10 billion
$J/\psi$ events collected with the BESIII detector. We find no
evidence of a signal and set upper limits at the $95\%$
confidence level on the product branching fraction $\mathcal{B}(J/\psi
\to \gamma a) \times \mathcal{B}(a \to \gamma \gamma)$ and the
axion-like particle photon coupling constant $g_{a \gamma \gamma}$ in
the ranges of $(3.7-48.5) \times 10^{-8}$ and $(2.2 -101.8)\times
10^{-4}$ GeV$^{-1}$, respectively, for $0.18 \le m_a \le 2.85~
\gevcc$. These are the most stringent limits to date in this mass
region.

\end{abstract}

\maketitle

Axions are hypothetical pseudoscalar gauge bosons originally proposed by Peccei and Quinn as a mechanism of spontaneous  U(1) symmetry breaking to resolve the strong $CP$ problem in quantum chromodynamics~\cite{Peccei, Weinberg, Wilczek} and later applied to the electroweak hierarchy problem~\cite{Graham}. Axion-like particles (ALPs) have the same quantum numbers as the QCD axion, but they have different masses and couplings. ALPs appear in various extensions of the Standard Model (SM), such as supersymmetry~\cite{Frere, Nelson, Bagger}, extended Higgs sector~\cite{Higgs-setr}, and string theory~\cite{Witten, Conlon, Svrcek, Arvanitaki}. The ALPs can act as a mediator between the dark sector and ordinary matter~\cite{Nomura, Freytsis}, and they could be cold dark matter candidates in certain situations~\cite{Preskill, Abbott, Dine, Battye}. A light scalar decaying to two photons could also be a signature of a $CP$-violating MSSM~\cite{Hesselbach} or Fermi-phobic Higgs bosons~\cite{Delgado}. Experimental constraints on these scenarios are important to understand the properties of the SM Higgs boson~\cite{Exotic}. In a simple scenario, the ALP, $a$, predominantly couples to a photon pair with a coupling constant $g_{a \gamma \gamma}$.  Ref.~\cite{graviton} has recently discussed the phenomenology of graviton-like spin-2 particles and the reinterpretation of existing experimental bounds on $g_{a \gamma \gamma}$ to set bounds on these possibilities.

 Experimental bounds on $g_{a \gamma \gamma}$ for masses of $a$ in the sub-MeV/$c^2$ range are available from laser experiments, solar photon instruments and astrophysical observations~\cite{Graham2, Cadamuro}, while the experimental bounds on the photon coupling in the MeV/$c^2$$-$GeV/$c^2$ range mainly come from beam-dump and high-energy collider experiments~\cite{Aloni, Dobrich, Banerjee}.

Limits on long-lived ALPs are derived from radiative decays $V \to \gamma + \rm{invisible}$ ($V=\Upsilon(1S), J/\psi$)~\cite{bes3higgs, bellehiggs} and the $e^+e^- \to \gamma + \rm{invisible}$ reaction~\cite{Dolan,BaBarinv, BaBarinv1, BaBarinv2}. On the other hand, short-lived ALPs decaying to a photon pair are searched for in $e^+e^- \to \gamma \gamma(\gamma)$ reactions~\cite{Knapen, Acciarri, Abudinen}, the light-by-light scattering process $\gamma \gamma \to \gamma \gamma$~\cite{Sirunyan, Aad}, $pp$ collisions~\cite{Enterria, Atlas, Atlas2} and radiative $J/\psi \to \gamma a$ decays, based on 2.7 billion $\psi(3686)$ events at BESIII~\cite{bes3ALP}.  The previous BESIII measurement~\cite{bes3ALP} selected a sample of approximately 1 billion $J/\psi$ events by tagging the pion-pair from $\psi(3686) \to \pi^+\pi^- J/\psi$ transition.  The ALP-photon coupling in the ALP mass range of (0.1-5) $\gevcc$~ is less constrained compared to the other mass regions. The best upper bound on $g_{a\gamma \gamma}$ mainly comes from the previous BESIII measurement~\cite{bes3ALP} for $0.165 \le m_a \le 1.468$ \gevcc~ and OPAL~\cite{Knapen} measurement in the rest of the mass region.  The large data sample collected by the BESIII detector at the $J/\psi$ resonance~\cite{njpsi} can further improve the sensitivity on $g_{a \gamma \gamma}$ in the mass region $0.18 \le m_a \le 2.85 ~\gevcc$~\cite{Jaeckel}.

In $J/\psi$ decays, ALP production predominantly proceeds via the radiative decay $J/\psi \to \gamma a$~\cite{Merlo}. The branching fraction of $J/\psi \to \gamma a$ relates to $g_{a\gamma \gamma}$ as~\cite{Merlo}

\begin{equation}
\frac{\mathcal{B}(J/\psi \to \gamma a)}{\mathcal{B}(J/\psi \to e^+e^-)}=\frac{m_{J/\psi}^2}{32\pi\alpha}g_{a\gamma\gamma}^2\left(1-\frac{m_a^2}{m_{J/\psi}^2} \right)^3, 
\label{reson}
\end{equation}

\noindent where $\alpha$ is the fine-structure constant, $m_a$ is the mass of the ALP and $m_{J/\psi}$ is the mass of the $J/\psi$ meson. Additional processes contributing to the same final state are the ALP-strahlung  $e^+e^- \to \gamma a$ process~\cite{Merlo} and the radiative production of pseudoscalar mesons through $J/\psi \to \gamma P$ with $P=\{\pi^0, \eta,\eta', \eta_c\}$~\cite{pdg, bes3jpstogp}.  

 This paper describes the search for a light pseudoscalar ALP in radiative decays of the $J/\psi$ using  $(1.0087 \pm 0.0044) \times 10^{10}$  $J/\psi$ events collected with the BESIII detector~\cite{njpsi}. This search assumes that the direct coupling of the ALP to charm quarks is negligible, and that the ALP produced in $J/\psi \to \gamma a$ couples to the virtual photon created in $c\bar{c}$ annihilation~\cite{Merlo}.  Since the ALP-strahlung process $e^+e^- \to \gamma a$ is indistinguishable from ALP production in radiative $J/\psi$ decays in the case of a null result, the expected $4.4\%$ ALP-strahlung contribution~\cite{Merlo} is subtracted from the experimental measurement to derive constraints on the $J/\psi  \to \gamma a$ branching ratio.  The interference between the ALP-strahlung process $e^+e^- \to \gamma a$ and radiative $J/\psi \to \gamma a$ is predicted to be negligible due to the small $J/\psi$ width~\cite{Merlo}.

The BESIII detector~\cite{bes3-det} records symmetric $e^+e^-$ collisions 
provided by the BEPCII storage ring~\cite{Yu:IPAC2016-TUYA01}
in the center-of-mass energy range from 2.0 to  4.95~GeV
with a peak luminosity of $1 \times 10^{33}\;\text{cm}^{-2}\text{s}^{-1}$ 
achieved at $\sqrt{s} = 3.77\;\text{GeV}$. 
BESIII has collected large data samples in this energy region~\cite{Ablikim:2019hff, JLu, JZhang}.  The cylindrical core of the BESIII detector covers 93\% of the full solid angle and consists of a helium-based
 multi-layer drift chamber~(MDC), a plastic scintillator time-of-flight
system~(TOF), and a CsI(Tl) electromagnetic calorimeter~(EMC),
which are all enclosed in a superconducting solenoidal magnet
providing a 1.0~T (0.9~T in 2012, for about 11\% of the dataset) 
magnetic field.  
The solenoid is supported by an
octagonal flux-return yoke with resistive plate counter muon
identification modules interleaved with steel. 
The charged-particle momentum resolution at $1~{\rm GeV}/c$ is
$0.5\%$. The EMC measures photon energies with a
resolution of $2.5\%$ ($5\%$) at $1$~GeV in the barrel (end cap)
region. The time resolution in the TOF barrel region is 68~ps, while
that in the end cap region is 110~ps.  The end cap TOF
system was upgraded in 2015 using multigap resistive plate chamber
technology, providing a time resolution of 60~ps. 
About 87\% of the dataset benefits from this upgrade.  

Monte Carlo (MC) simulated data samples produced with a {\sc Geant4}-based software package~\cite{geant4}, which includes the geometrical acceptance of the detector and time-dependent beam related backgrounds, are used to optimize the event selection criteria, study the potential backgrounds, and evaluate the reconstruction efficiency. A MC sample of 10 billion inclusive $J/\psi$ decays is used for background studies with the TopoAna tool~\cite{topoana}. In this sample, the known decay modes of $J/\psi$ are simulated by {\sc EvtGen}~\cite{evtgen} with their branching fractions taken from the Particle Data Group (PDG)~\cite{pdg} and the remaining unknown decay modes by {\sc Lundcharm}~\cite{Lundcharm}. The production of the $J/\psi$ resonance via $e^+e^-$ annihilation, including beam energy spread and initial-state radiation (ISR), is simulated by {\sc KKMC}~\cite{kkmc}. The backgrounds from the SM processes $e^+e^- \to \gamma \gamma (\gamma)$ and $e^+e^- \to \gamma P$  are estimated with 166.3~pb$^{-1}$ of continuum data collected at the center-of-mass energy of 3.08 GeV~\cite{njpsi}. To evaluate the signal efficiency, simulated signal MC events are generated for 33 values of $m_a$ ranging from $0.1$ \gevcc~ to $3.0$ \gevcc~ with a phase-space model for $a \to \gamma \gamma$ and a $P$-wave model for the $J/\psi \to \gamma a$ decay~\cite{evtgen}. To avoid any bias, a semi-blind procedure is adopted in which approximately $10\%$ of full $J/\psi$ data is used to check the agreement between data and background predictions and validate the fitting approach. The rest of the $J/\psi$ dataset is blinded until the analysis procedure is frozen.

We select the events of interest with at least three photon candidates and zero charged tracks in the final state. The photon candidates are reconstructed using the energy clusters deposited in the EMC barrel region with a minimum energy of 25 MeV. Endcap clusters are not used in order to suppress the background from $e^+e^- \to \gamma \gamma(\gamma)$ events. The energy deposited in the nearby TOF scintillators is also included to improve the reconstruction efficiency and energy resolution.  The EMC time difference between any two photons is required to be within the range of $-500 < \Delta T < 500$ ns to suppress electronics noise and energy deposits unrelated to the events.  
To improve the mass resolution,
a four-constraint (4C) kinematic fit is performed, constraining the mass of the three-photon system to the center-of-mass energy. If there is more than one $J/\psi \to \gamma \gamma \gamma$ combination, the candidate with the minimum value of the $\chi^2$ from the kinematic fit ($\chi_{\rm 4C}^2$) is retained. The $\chi_{\rm 4C}^2$ is required to be less than 30. A similar 4C kinematic fit is also separately performed with all possible two, four and five photon candidate hypotheses. We require that the $\chi_{\rm 4C}^2$ of the three photons candidate hypothesis is less than that for the other hypotheses to avoid any wrong combination of soft photons. The requirements of $\chi_{\rm 4C}^2 < \chi_{\rm 4C}^2(2 \gamma)$ and $\chi_{\rm 4C}^2 < \chi_{\rm 4C}^2 (n \gamma)$ ($n=4, 5$) suppress the backgrounds from $e^+e^- \to \gamma \gamma$ process and $J/\psi \to \gamma \pi^0 \pi^0$ decays when a low-energy photon escapes undetected.  The selected three photon candidates are used for further analysis.  

In order to further suppress the $e^+e^- \to \gamma \gamma(\gamma)$ events
with low-energy EMC clusters,
the energy difference between the first (second) and third photon candidates, $\Delta E_{\gamma_{ij}}$, is required to be less than $1.46~ (1.41)~ \gev$, where the first, second and third photons are sorted by decreasing energy. The absolute azimuthal angle difference between the third and first photon candidates, $\Delta \phi_{\gamma_{31}}$, is also required to be greater than 1 radian. 

 At this stage, the dominant background contributions come mainly from $J/\psi \to \gamma P$, where $P$ denotes the pseudoscalar mesons $\pi^0, \eta, \eta', \eta_c$, as also seen in Refs.~\cite{bes3ALP,bes3jpstogp}. To validate the signal extraction procedure (see below) the same fit is used to extract the branching fractions of $J/\psi \to \gamma (\pi^0,\eta,\eta') \to \gamma \gamma \gamma$. 
After accounting for the peaking background contributions, such as $J/\psi \to \gamma \pi^0\pi^0$ in $J/\psi \to \gamma \pi^0$ decay as described in Ref.~\cite{bes3jpstogp}, the results are compatible with the measurements of Ref.~\cite{bes3jpstogp} and PDG averages~\cite{pdg} within their uncertainties. 
We further reject the events where the di-photon invariant mass ($m_{\gamma \gamma}$) for any photon combination is in the regions
$0.11 \le m_{\gamma \gamma} \le 0.16$ GeV/$c^2$, $0.52 \le m_{\gamma \gamma} \le 0.56$ GeV/$c^2$, $0.92 \le m_{\gamma \gamma} \le 0.99$ GeV/$c^2$, and $2.92 \le m_{\gamma \gamma} \le 3.04$ GeV/$c^2$ to suppress the backgrounds from $J/\psi \to \gamma \pi^0$, $J/\psi \to \gamma \eta$, $J/\psi \to \gamma \eta'$ and $J/\psi \to \gamma \eta_c$, respectively.

After applying the above selection criteria, the signal yields are extracted from the data by performing a series of one-dimensional unbinned extended maximum likelihood (ML) fit to the $m_{\gamma \gamma }$ distribution, which includes all combinations of two photon pairs. The fit function includes the contributions of signal and  backgrounds and is described in more detail below. 

\begin{figure}
  \centering
 \includegraphics[width=0.50\textwidth]{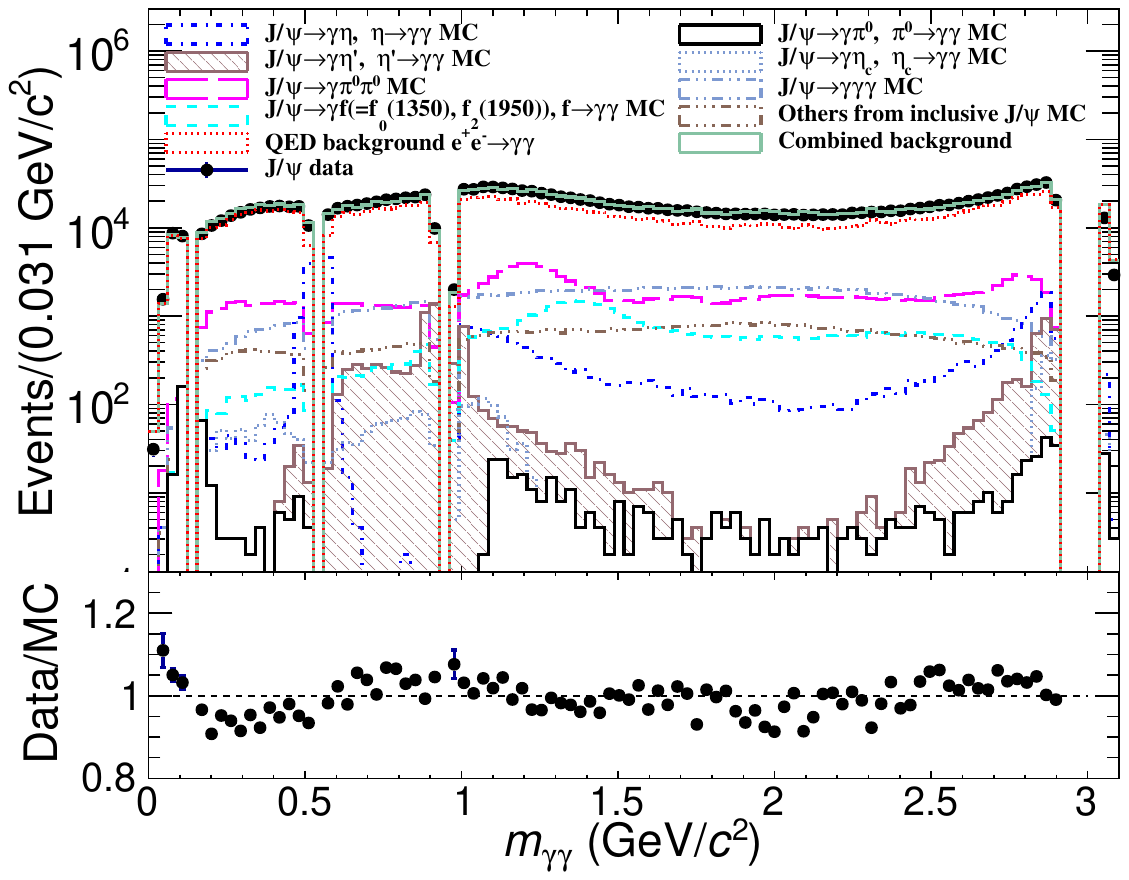}
  
  \caption{Distribution of the di-photon invariant mass, $m_{\gamma \gamma}$, for data (black dots with error bars) together with the background predictions of the QED $e^+e^- \to \gamma \gamma$ process from data collected at 3.08 GeV (dotted red curve), $J/\psi \to \gamma \pi^0\pi^0$ (long-dashed pink curve), $J/\psi \to \gamma \gamma \gamma$ (double-dotted long-dashed grey curve), $J/\psi \to \gamma f$ $(f=f_0(1350), f_2(1950))$ (long-dashed cyan curve), $J/\psi \to \gamma P$ (brown pattern, dashed-dotted blue, black and dotted grey histograms) and other backgrounds from inclusive $J/\psi$ decays (triple dotted long-dashed brown curve). The solid green olive histogram represents the combined background. The bottom figure shows the ratio of data to the combined background predictions.  }
  \label{mgg}
\end{figure}

 Fig.~\ref{mgg} shows the $m_{\gamma \gamma }$ distribution
for data and various background predictions. The dominant background comes from the QED process $e^+e^- \to \gamma \gamma (\gamma)$, which is predicted from the continuum data collected at 3.08 GeV. The $m_{\gamma \gamma}$ distribution for data is generally well-described by the background predictions, except in the low-mass region, where {\sc KKMC}~\cite{kkmc} fails to reproduce the events for $J/\psi$ decaying to multi-photons in the final state. This disagreement has a minor impact on the ALP search because the signal extraction procedure does not depend on the predictions of the background yields. In order to mitigate the effect from the tails of the peaking backgrounds $J/\psi \to \gamma P$, the fit is performed in different $m_{\gamma \gamma}$ intervals for various $m_a$ points, as described in Table~\ref{fit-interval}. 

\begin{table}
\centering
\caption{The fit intervals  of  $m_{\gamma \gamma}$ for various $m_{a}$ points. }
\begin{tabular}{c|c|c}
        \hline  

 $m_a$ range   &   $m_{\gamma \gamma}$ fit interval & Polynomial  \\
 (GeV/$c^2$)   &   (GeV/$c^2$)                      & function order \\
        \hline  
 $0.180 - 0.420$    & $0.16, 0.46$  & $4^{\rm th}$ \\
 $0.421 - 0.490$    & $0.39, 0.51$  & $5^{\rm th}$ \\
 $0.610 - 0.880$    & $0.59, 0.90$  & $5^{\rm th}$ \\
 $1.020 - 1.099$    & $1.00, 1.20$  & $5^{\rm th}$ \\
 $1.100 - 2.770$    & $m_a-0.10, m_a+0.10$   & $3^{\rm rd}$ \\
 $2.772 - 2.850$    & $2.70, 2.88$  & $4^{\rm th}$ \\   \hline 
\end{tabular}
\label{fit-interval}
\end{table}

Simulated data samples are used to construct the signal and  background probability density functions (PDFs). The signal PDF is described by the sum of two Crystal Ball functions with common mean and opposite-side tails. The efficiency and parameters of the Crystal Ball functions are obtained from the simulated signal MC samples by performing a fit to the $m_{\gamma \gamma}$ distribution. This fit includes the contributions of signal and combinatorial background from wrong $\gamma\gamma$ combinations, described by a first-order Chebyshev function. The $m_a$ resolution varies from $6~\mevcc$ to $15~\mevcc$ while the signal selection efficiency, $\epsilon$, varies from $28.2\%$ to $39.0\%$ depending on the $a$ mass. The PDF parameters and efficiency of the signal are interpolated linearly between the mass points of the generated signal MC samples. The background PDF is described by $3^{\rm rd}$-, $4^{\rm th}-$ and $5^{\rm th}$-order Chebyshev functions in various $m_{\gamma \gamma}$ fit regions detailed in Table~\ref{fit-interval}. 

The search is performed in $1.0~\mevcc$ steps in the mass range of $0.18 \le m_a \le 1.5~\gevcc$ and $2.0~\mevcc$ steps for higher $m_a$ values. The free parameters of the fit are the number of signal events ($N_{\rm sig}^{\rm dat}$), which includes the contributions of both radiative $J/\psi \to \gamma a$ and ALP-strahlung $e^+e^- \to \gamma a$ process, the number of background events, and the shape parameters of the background PDF. 
To take into account a possible systematic uncertainty associated with the choice of PDFs, 
the fit is repeated with an alternative signal PDF described by a Cruijff function~\cite{Cruijff}, 
 or increasing by one the order of the Chebyshev polynomial function
for the background PDF at each $m_a$ point.
The fit with the largest signal yield, giving the worst upper limit, is chosen to produce the final result. The final signal yield $N_{\rm sig}$ is  $95.6\%$ of the  $N_{\rm sig}^{\rm dat}$ after subtracting the contribution from the ALP-strahlung process~\cite{Merlo}. 

 We calculate the product branching fraction of $J/\psi \to \gamma a$ and $a \to \gamma \gamma$ at each $m_a$ point using the following formula,

\begin{equation}
\mathcal{B} (J/\psi \to \gamma a) \times \mathcal{B}(a \to \gamma \gamma)=\frac{N_{\rm sig}}{\epsilon N_{J/\psi}}.
\label{bf}
\end{equation} 

\begin{figure} \centering
  \includegraphics[width=0.5\textwidth]{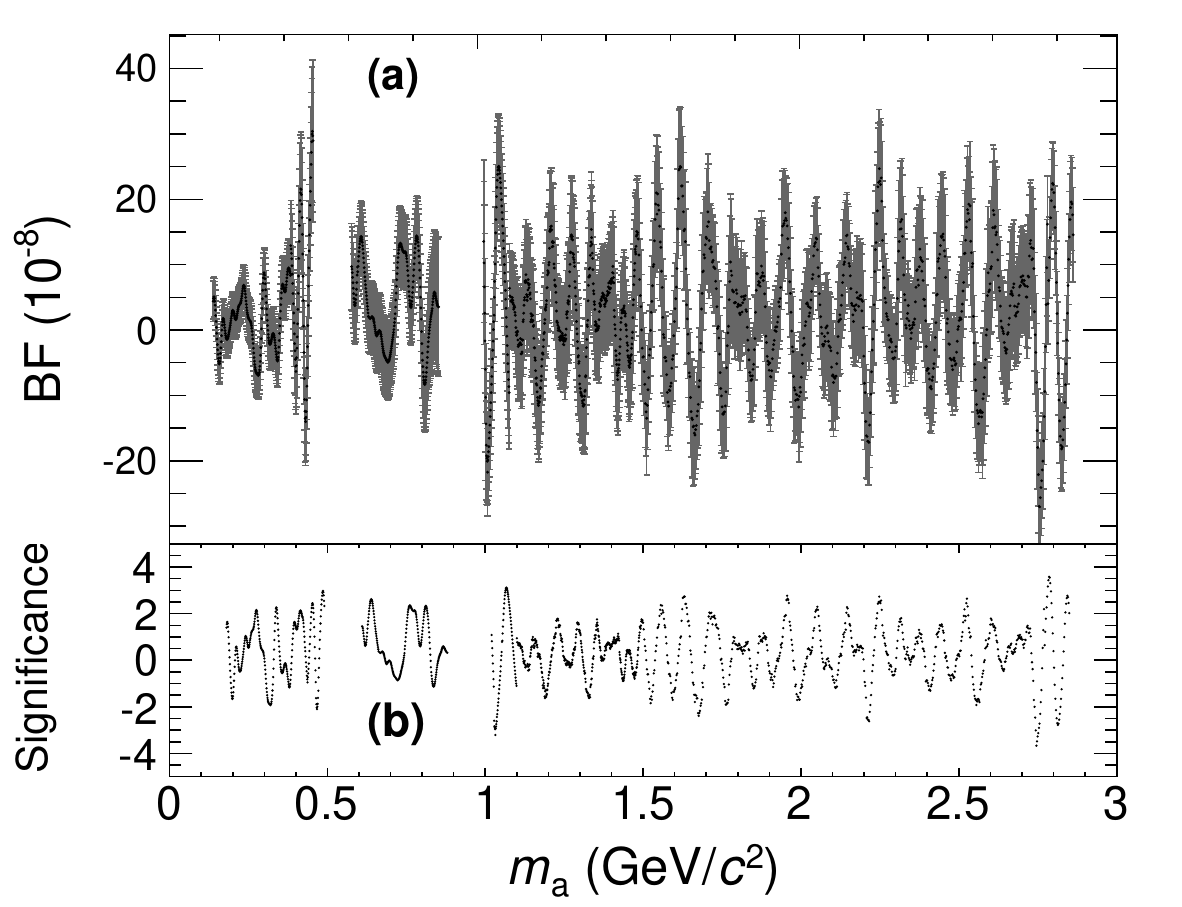} \caption{
    (a) Product branching fraction $\mathcal{B}(J/\psi \to \gamma a) \times
    \mathcal{B}(a \to \gamma \gamma)$ (BF) and (b) the signal significance
     versus $m_a$ obtained from the fit, as described in the text.}
  \label{yield}
\end{figure}

\begin{figure}
  \centering
 \includegraphics[width=0.50\textwidth]{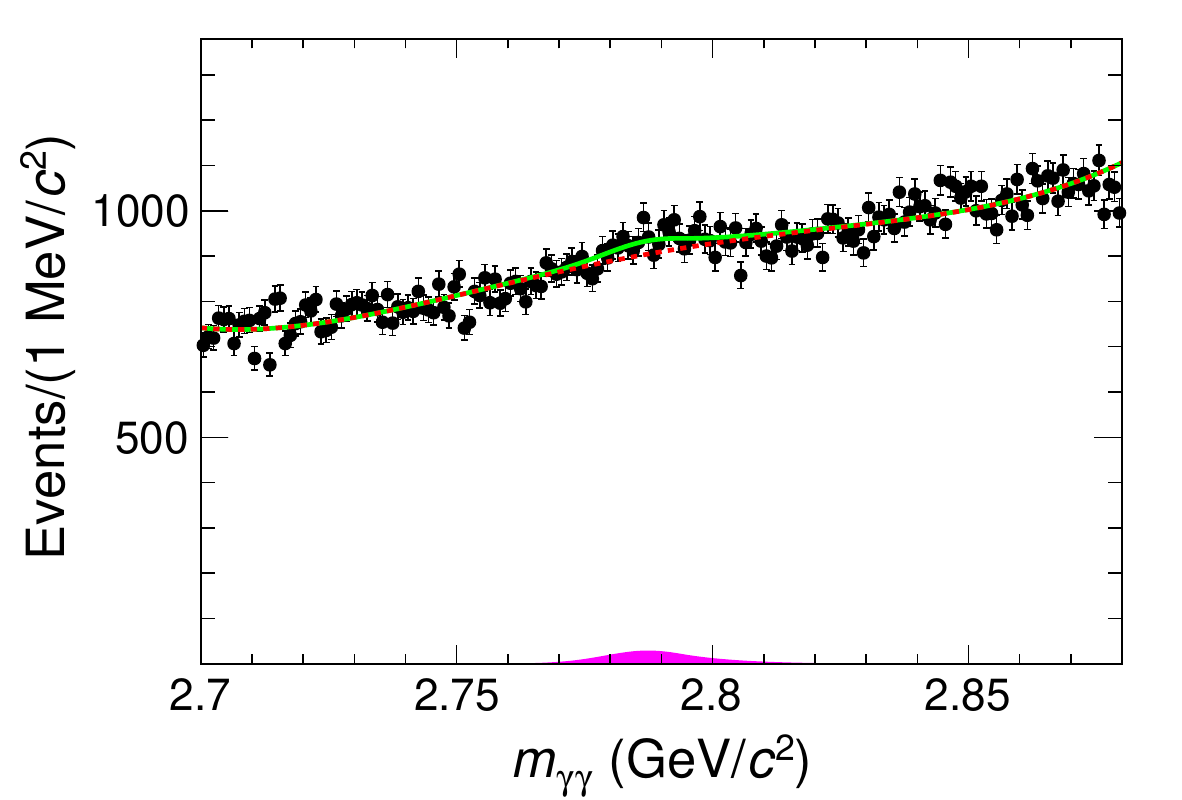}
  
  \caption{Fit to the $m_{\gamma \gamma}$ distribution for $m_a=2.786$ GeV/$c^2$. The black dots with error bars represent data, the dashed red curve is the continuum background, the filled pink curve is the signal PDF, and the solid green curve is the total fit result. }
  \label{mgg_proj}
\end{figure}

\noindent Figure~\ref{yield} shows the product branching fraction $\mathcal{B} (J/\psi \to \gamma a) \times \mathcal{B}(a \to \gamma \gamma)$ and the statistical significance, defined as $\mathcal{S} =\sqrt{-2{\rm ln}(\mathcal{L}_{0}/\mathcal{L}_{\rm max}})$, as a function of $m_a$. In this formula $\mathcal{L}_{\rm max}$ and $\mathcal{L}_0$ are the likelihood functions for the number of signal events obtained from the fit and fixed at zero, respectively. The largest value of upward local significance is determined to be $3.5\sigma$ at $m_a=2.786$ GeV/$c^2$. The fit result for the corresponding $m_a$ point is shown in Fig.~\ref{mgg_proj}. We estimate the probability of observing a fluctuation of $\mathcal{S} \ge 3.5 \sigma$ to be $5.2\%$ using a large ensemble of pseudo-experiments~\cite{trial-factor}. The corresponding global significance value is $1.6 \sigma$. Thus, we conclude that no significant evidence of signal events is found within the investigated $m_a$ regions.

As follows from Eqs.~\ref{reson} and ~\ref{bf}, the systematic uncertainties for the coupling of the ALP to a photon pair and the branching fraction measurements include those from the number of signal events, the reconstruction efficiency, the total number of $J/\psi$ events, and the branching fraction of $J/\psi \to e^+e^-$. The uncertainties associated with the number of signal events are additive because they originate from the PDF parameters of signal and backgrounds and the fit bias. The additive systematic uncertainty affects the significance of any observation and does not scale with the number of reconstructed signal events. We take into account the uncertainties associated with the PDF parameters of signal and  backgrounds by performing the alternative fits at each $m_a$ point using alternative PDFs, as described above. We utilize a large number of pseudo-experiments to test the reliability of the ML fit procedure and the accuracy of the signal and background modeling. The signal yields are extracted from these samples using the same fit method described above. The biases are consistent with zero and their average uncertainty is 9.2 events, which is taken as an additional additive systematic uncertainty ($\sigma_{\rm add}$).

The other sources of the systematic uncertainties are multiplicative; they scale with the number of reconstructed events and do not affect the significance of a yield. The uncertainty associated with the reconstruction efficiency includes contributions from the 4C kinematic fit, the selection criteria of $\Delta E_{ij}$ and $\Delta \phi_{13}$ and photon detection efficiency.

We use a control sample of $J/\psi \to \gamma \eta$, $\eta \to \gamma \gamma$ to evaluate the systematic uncertainties associated with the $\chi_{\rm 4C}^2$, $\Delta E_{\gamma_{13}}$, $\Delta E_{\gamma_{23}}$, $J/\psi \to \gamma P$ veto, and $\Delta \phi_{13}$ requirements. The signal yields for $J/\psi \to \gamma \eta$ from both data and MC simulation are extracted by performing an ML fit to the $m_{\gamma \gamma}$ distribution using the same fit procedure as described above. The corresponding systematic uncertainties, computed as the relative change in efficiency between data and MC simulation, are determined to be $2.3\%$, $0.1\%$, $0.1\%$, $0.8\%$ and $0.1\%$, respectively. The systematic uncertainty associated with the total number of $J/\psi$ events is determined to be $0.44\%$ with a sample of inclusive $J/\psi$ hadronic events~\cite{njpsi}.

A control sample of $e^+e^- \to \gamma \mu^+\mu^-$ is used to evaluate the systematic uncertainty associated with the photon reconstruction efficiency. An ISR photon in this control sample is predicted using the four momenta of the two charged particles. This sample also includes the contributions from $e^+e^- \to \gamma \pi^+\pi^-$ and $J/\psi \to \gamma \pi^+ \pi^-$, including all the possible intermediate resonances. The relative difference of efficiency between data and MC simulation is determined to be $0.2\%$ per photon ~\cite{photon-detection}. Thus, the total systematic uncertainty due to photon reconstruction efficiency for three photon candidates is $0.6\%$. The uncertainty associated with the branching fraction of $J/\psi \to e^+e^-$ is $0.5\%$ taken from the PDG~\cite{pdg}. The total multiplicative systematic uncertainty ($\sigma_{\rm mult}$) is obtained by adding the individual ones in quadrature, and is determined to be $2.6\%$  for both a product branching fraction $\mathcal{B}(J/\psi \to \gamma a) \times \mathcal{B}(a \to \gamma \gamma)$ and a coupling of ALP to a photon pair $g_{a\gamma \gamma}$. All sources of systematic uncertainties are summarized in Table~\ref{syst}. We calculate the final systematic uncertainty as $\sqrt{\sigma_{\rm add}^2 + (\sigma_{\rm mult} \times N_{\rm sig})^2}$.

\begin{table}[htb]
\centering 
  \caption{The sources of systematic uncertainties. The uncertainties associated with the signal and background PDFs are incorporated in the fit, 
as described in the text.}

\begin{tabular}{c |c|c}
  \hline
 Source  & \multicolumn{2}{|c}{ Uncertainty} \\ \hline
        & $J/\psi \to \gamma  a$ & $g_{a \gamma \gamma}$ (GeV)$^{-1}$ \\
\hline \hline
\multicolumn{3}{l}{{\hskip 0.8cm} Additive (events) } \\
\hline
 Fit Bias                     & 9.2   & 9.2             \\
\hline
\hline
\multicolumn{3}{l}{{\hskip 0.7cm} Multiplicative ($\%$)} \\
\hline
$\chi_{4C}^2$                          & 2.3 & 2.3  \\
$\Delta E_{\gamma_{13}}$               & 0.1 & 0.1  \\
$\Delta E_{\gamma_{23}}$               & 0.1 & 0.1  \\
$J/\psi \to \gamma P$ veto       & 0.8 & 0.8        \\
$\Delta\phi_{\gamma_{31}}$             & 0.1 & 0.1  \\
$\mathcal{B}(J/\psi \to e^+e^-)$       & --  & 0.5  \\
$J/\psi$ counting                      &0.44 & 0.44 \\
Photon detection efficiency            & 0.6 & 0.6  \\ \hline \hline
Total                                  &2.6  & 2.6  \\ \hline

\hline
\end{tabular}
\label{syst}
\end{table}

\begin{figure}
  \centering
 \includegraphics[width=0.50\textwidth]{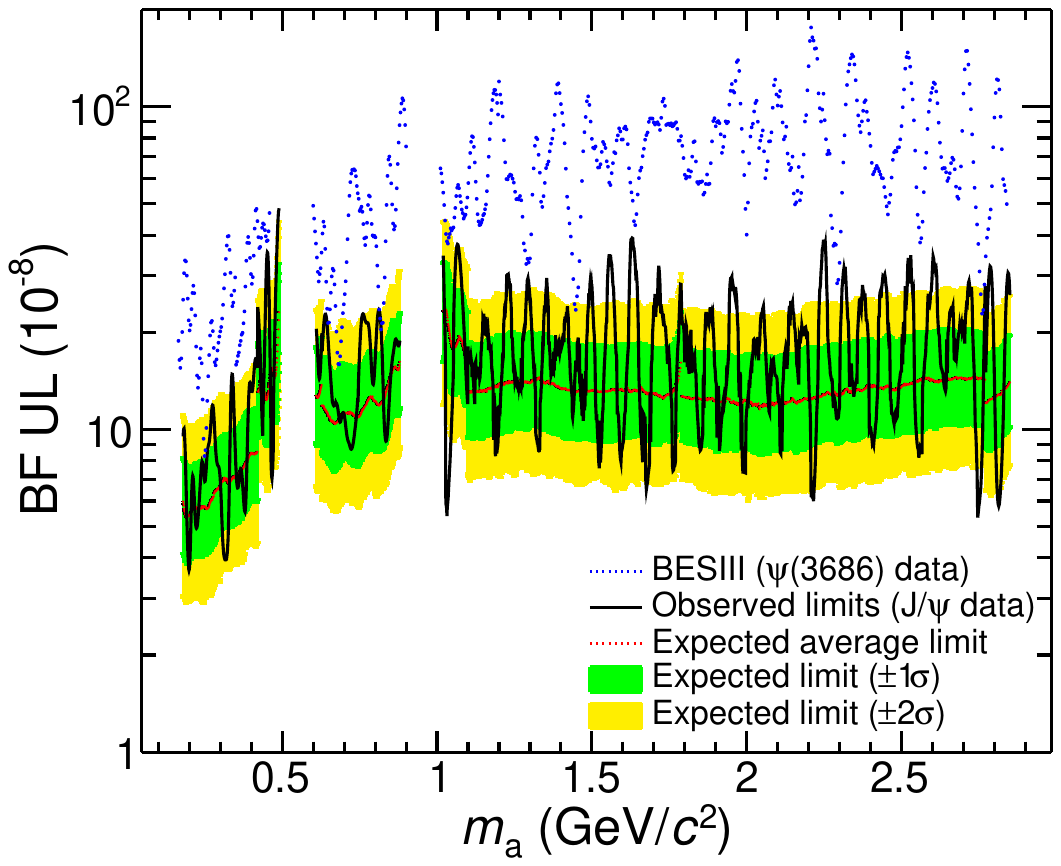}
  
  \caption{Upper limits at the $95\%$ C.L.\ on the product branching fraction $\mathcal{B}(J/\psi \to \gamma a) \times \mathcal{B}(a \to \gamma \gamma)$,  including the systematic uncertainties, and compared with the previous BESIII measurements~\cite{bes3ALP} as a function of $m_a$. The expected limit bands at $\pm 1\sigma$ and $\pm 2 \sigma$ levels,  obtained from a large ensemble of pseudo-experiments, include statistical uncertainty only.  }
  \label{BFUL}
\end{figure}

\begin{figure}
  \centering
 \includegraphics[width=0.50\textwidth]{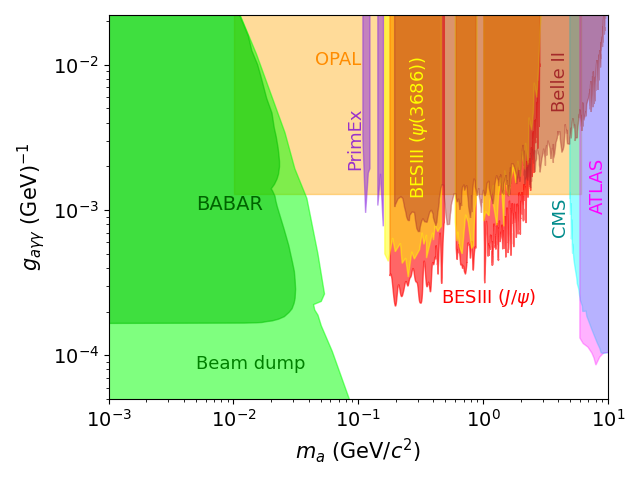}
  
  \caption{The $95\%$ CL upper limits on the coupling of ALP to a photon pair ($g_{a \gamma \gamma}$)  together with BaBar~\cite{Dolan}, CMS~\cite{Sirunyan}, ATLAS~\cite{Aad}, PrimEx~\cite{Aloni}, beam-dump~\cite{Dobrich, Banerjee},  OPAL~\cite{Knapen}, previous BESIII~\cite{bes3ALP}  and Belle II~\cite{Abudinen} measurements as a function of $m_a$.}
  \label{gagg}
\end{figure}

Since no significant evidence of any signal events is found, we set the $95\%$ confidence level (CL) upper limits on the product branching fraction $\mathcal{B}(J/\psi \to \gamma a) \times \mathcal{B}(a \to \gamma \gamma)$ as a function of $m_a$ using a Bayesian approach  with a uniform prior. The systematic uncertainty is included by convolving the likelihood function with a Gaussian function having a width equal to the systematic uncertainty. The limits range between $(3.7-48.5) \times 10^{-8}$ for $0.18 \le m_a \le 2.85$ GeV/$c^2$, as shown in Fig.~\ref{BFUL} together with expected limit bands at $\pm 1 \sigma$ and $\pm 2 \sigma$ levels obtained from a large ensemble of pseudo-experiments. Our results improve upon the previous  BESIII measurement~\cite{bes3ALP}   by an average factor of $8-9$. 
Finally, we also calculate the $95\%$ CL upper limit on $g_{a \gamma \gamma}$ using Eq.~\ref{reson} while assuming $\mathcal{B}(a \to \gamma \gamma)=1$ and including the additional sources of the systematic uncertainties on $\mathcal{B}(J/\psi \to e^+e^-)$  using a Bayesian approach  with a uniform prior in a negative log likelihood versus  $g_{a\gamma \gamma}^2$ curve. The corresponding exclusion range is shown in Fig.~\ref{gagg}. The limit varies within $(2.2 - 101.8) \times 10^{-4}$  (GeV)$^{-1}$ for $0.18 \le m_a \le 2.85~\gevcc$ and  improves previous bounds set by  previous BESIII~\cite{bes3ALP} and Belle II~\cite{Abudinen} by a factor of about  3 and 5, respectively.  It has also two fold improvement over the OPAL measurement~\cite{Knapen} for $1.468 \le m_a \le 2.2$ \gevcc.

In summary, we search for di-photon decays of a light pseudoscalar particle in radiative $J/\psi$ decays, using 10 billion $J/\psi$ events collected by the BESIII detector. No significant evidence of signal events is found, and we set $95\%$ CL upper limits on the product branching fraction $\mathcal{B}(J/\psi \to \gamma a) \times \mathcal{B}(a \to \gamma \gamma)$ and the coupling of the ALP to a photon pair $g_{a \gamma \gamma}$ as a function of $m_a$. The corresponding limits are more stringent than the existing limits from OPAL~\cite{Knapen}, previous BESIII~\cite{bes3ALP} and Belle II~\cite{Abudinen}  for $0.18 \le m_a \le 2.85$ GeV/$c^2$. These improved limits can significantly constrain the parameter spaces of the extended Higgs sector models~\cite{Higgs-setr, Merlo, Hesselbach, Delgado} and other new physics models~\cite{graviton, Frere, Nelson, Bagger, Freytsis} in the investigated mass region.


\textbf{Acknowledgement}

The BESIII Collaboration thanks the staff of BEPCII and the IHEP computing center for their strong support. This work is supported in part by National Key R\&D Program of China under Contracts Nos. 2020YFA0406300, 2020YFA0406400; National Natural Science Foundation of China (NSFC) under Contracts Nos. 11635010, 11735014, 11835012, 11935015, 11935016, 11935018, 11961141012, 12025502, 12035009, 12035013, 12061131003, 12192260, 12192261, 12192262, 12192263, 12192264, 12192265, 12221005, 12225509, 12235017, 11950410506, 11705192; the Chinese Academy of Sciences (CAS) Large-Scale Scientific Facility Program; the CAS Center for Excellence in Particle Physics (CCEPP); Joint Large-Scale Scientific Facility Funds of the NSFC and CAS under Contract No. U1832207; CAS Key Research Program of Frontier Sciences under Contracts Nos. QYZDJ-SSW-SLH003, QYZDJ-SSW-SLH040; 100 Talents Program of CAS; The Institute of Nuclear and Particle Physics (INPAC) and Shanghai Key Laboratory for Particle Physics and Cosmology; European Union's Horizon 2020 research and innovation programme under Marie Sklodowska-Curie grant agreement under Contract No. 894790; German Research Foundation DFG under Contracts Nos. 455635585, Collaborative Research Center CRC 1044, FOR5327, GRK 2149; Istituto Nazionale di Fisica Nucleare, Italy; Ministry of Development of Turkey under Contract No. DPT2006K-120470; National Research Foundation of Korea under Contract No. NRF-2022R1A2C1092335; National Science and Technology fund of Mongolia; National Science Research and Innovation Fund (NSRF) via the Program Management Unit for Human Resources \& Institutional Development, Research and Innovation of Thailand under Contract No. B16F640076; Polish National Science Centre under Contract No. 2019/35/O/ST2/02907; The Swedish Research Council; U. S. Department of Energy under Contract No. DE-FG02-05ER41374


\begin{thebibliography}{99}
\bibitem{Cruijff}  The Cruijff function is an asymmetric Gaussian function having a common mean ($\mu$),  different left-right resolutions ($\sigma_{L,R}$) and non-Gaussian tails ($\alpha_{L,R}$): $f(x)={\rm exp}((x-\mu)^2/(2\sigma_{L,R}^2+\alpha_{L,R}(x-\mu)^2))$.
\end{thebibliography}
\end{document}